\documentclass{aa}  

\usepackage{amsmath,amssymb}
\usepackage[breaklinks=true]{hyperref}

\usepackage{graphicx}
\usepackage{xcolor}
\usepackage[normalem]{ulem}
\usepackage{listings}
\definecolor{dkgreen}{rgb}{0,0.6,0}
\definecolor{gray}{rgb}{0.5,0.5,0.5}
\definecolor{mauve}{rgb}{0.58,0,0.82}
\usepackage{txfonts}
\include{notations}

\providecommand{\kms}{\ensuremath{\rm \,km\,s^{-1}}\xspace}
\providecommand{\masyr}{\ensuremath{\rm \,mas\,yr^{-1}}\xspace}

\providecommand{\kmskpc}{\ensuremath{\rm \,km\,s^{-1}\,kpc^{-1}\xspace}}
\providecommand{\lzunit}{\ensuremath{\rm \,km\,s^{-1}\,kpc\xspace}}

\providecommand{\kpc}{\ensuremath{\,\rm kpc}\xspace}

\providecommand{\kms}{\ensuremath{\textrm{km\,s}^{-1}}}

\providecommand{\masyr}{\ensuremath{\textrm{mas\,yr}^{-1}}}

\providecommand{\gaia}{\textit{Gaia}}

\newcommand\gdrtwo{\gaia~DR2}
\newcommand\gdrthree{\gaia~DR3}
\newcommand\gedrthree{\gaia~EDR3}

\newcommand{\vlos}{$V_{\rm los}$}
\newcommand{\vrad}{$V_R$}
\newcommand{\vtan}{$V_{\phi}$}
\newcommand{\vz}{$V_{Z}$}

\newcommand{\lz}{$L_{Z}$}
\newcommand{\lsun}{$L_{Z, \odot}$}
\newcommand{\jr}{$J_{R}$}
\newcommand{\jphi}{$J_{\phi}$}
\newcommand{\jz}{$J_{Z}$}

\newcommand{\galah}{{\sl GALAH}}

\newcommand{\apogee}{{\sl APOGEE}}
\newcommand{\rave}{{\sl RAVE}}
\newcommand{\lamost}{{\sl LAMOST}}

\newcommand{\rgb}{{\sl RGB}}
\newcommand{\rvs}{{\sl RVS}}
\newcommand{\agama}{{\sl AGAMA}}

\newcommand{\drimmelgaia}{{GD22}}
\newcommand{\paul}{{MC22}}
\newcommand{\acantoja}{{AC21}}

\newcommand{\cautunpot}{{\sl Cautun20}}
\newcommand{\mcmillanpot}{{\sl McMillan17}}

\newcommand{\bovypot}{{\sl MWPotential2014}}

\usepackage{graphicx}
\usepackage{txfonts}
\usepackage{hyperref}
\newcommand{\orcit}[1]{\protect\href{https://orcid.org/#1}{\protect\includegraphics[width=8pt]{orcid.png}}}

\makeatletter
\renewcommand*\maketitle{%
  \thispagestyle{firstpage}
\begingroup
    \if@wideboxfn
    \setlength\bibindent{1.4\parindent}
    \else
    \setlength\bibindent{\parindent}
    \fi
    \renewcommand*\thefootnote{\@fnsymbol\c@footnote}%
    \renewcommand\@makefntext[1]{%
    \ifaa@longfn\hsize\textwidth\fi
    \noindent
    \hb@xt@\bibindent{\hss\@makefnmark\enspace}##1}
  \ifaa@twocolumn
  \begingroup
    \begin{aa@strip}
          \aa@maketitle
    \end{aa@strip}
    \@thanks	  	
  \endgroup
  \else
    \begingroup
      \let\thanks\footnote
      \aa@maketitle
    \endgroup
  \fi
\endgroup
  \setcounter{footnote}{0}%
}
\makeatother
\makeatletter
\DeclareRobustCommand*{\fieldName}[1]{%
  \begingroup\@fieldName\scantokens{\texttt{\small {#1}}\noexpand}\endgroup}
\begingroup\lccode`\~=`\_\relax
   \lowercase{\endgroup\def\@fieldName{\catcode`\_=\active \let~\_}}
\makeatother

\providecommand{\linktodoc}{https://geapre.esac.esa.int/archive/documentation/GDR3}  

\begin{document}

   \title{A new resonance-like feature in the outer disc of the Milky Way
   }

\author{ 
Drimmel, R.\inst{1}  \and
Khanna, S.\inst{1} \and
D'Onghia, E.\inst{2} \and 
Tepper-Garc\'ia, T.\inst{3,4} \and
Bland-Hawthorn, J.\inst{3,4} \and
Chemin, L.\inst{5}  \and
Ripepi, V.\inst{6} \and
Romero-G\'omez, M.\inst{7,8,9}  \and
Ramos, P.\inst{7,8,9,10}  \and
Poggio, E.\inst{1,11}  \and
Andrae, R.\inst{12} \and
Blomme, R.\inst{13} \and
Cantat-Gaudin, T.\inst{12} \and
Castro-Ginard, A.\inst{14} \and
Clementini, G.\inst{15} \and
Figueras, F.\inst{7,8,9} \and
Fouesneau, M.\inst{12} \and
Fr\'emat, Y.\inst{13} \and
Lobel, A.\inst{13}  \and
Marshall, D.\inst{16} \and
Muraveva, T.\inst{15} 
}

\institute{INAF - Osservatorio Astrofisico di Torino, via Osservatorio 20, 10025 Pino Torinese (TO), Italy\\    \email{ronald.drimmel@inaf.it}
         \and
  Department of Astronomy, University of Wisconsin-Madison, 475 North Charter Street, Madison, WI 53706, USA
  \and
  Sydney Institute for Astronomy, School of Physics, University of Sydney, NSW 2006, Australia
  \and
  Centre of Excellence for All Sky Astrophysics in Three Dimensions (ASTRO-3D), Australia
  \and
  Centro de Astronom\'{i}a - CITEVA, Universidad de Antofagasta, Avenida Angamos 601, Antofagasta 1270300, Chile
  \and 
  INAF - Osservatorio Astronomico di Capodimonte, Via Moiariello 16, 80131, Napoli, Italy
  \and
  Institut de Ci\`{e}ncies del Cosmos (ICCUB), Universitat  de  Barcelona  (IEEC-UB), Mart\'{i} i  Franqu\`{e}s  1, 08028 Barcelona, Spain
  \and
  Departament de Física Quàntica i Astrofísica (FQA), Universitat de Barcelona (UB), C Martí i Franquès, 1, 08028 Barcelona, Spain
  \and
  Institut d’Estudis Espacials de Catalunya (IEEC), C Gran Capità, 2-4, 08034 Barcelona, Spain
  \and
  National Astronomical Observatory of Japan, Mitaka-shi, Tokyo 181-8588, Japan
  \and 
  Universit\'{e} C\^{o}te d'Azur, Observatoire de la C\^{o}te d'Azur, CNRS, Laboratoire Lagrange, Bd de l'Observatoire, CS 34229, 06304 Nice Cedex 4, France
  \and
  Max Planck Institute for Astronomy, K\"{ o}nigstuhl 17, 69117 Heidelberg, Germany
  \and
  Royal Observatory of Belgium, Ringlaan 3, 1180 Brussels, Belgium
  \and
  Leiden Observatory, Leiden University, Niels Bohrweg 2, 2333 CA Leiden, The Netherlands
  \and
  INAF - Osservatorio di Astrofisica e Scienza dello Spazio di Bologna, via Piero Gobetti 93/3, 40129 Bologna, Italy
  \and
  IRAP, Universit\'{e} de Toulouse, CNRS, UPS, CNES, 9 Av. colonel Roche, BP 44346, 31028 Toulouse Cedex 4, France
         }
         
       \date{Received ; accepted }

 
  \abstract{  
  Modern astrometric and spectroscopic surveys have revealed a wealth of structure in the phase space of stars in the Milky Way, with evidence of resonance features and non-equilibrium processes. Using \gaia's third data release, we present evidence of a new resonance-like feature in the outer disc of the Milky Way. The feature is most evident in the angular momentum distribution of the young Classical Cepheids, a population for which we can derive accurate distances over much of the Galactic disc. We then search for similar features in the outer disc using a much larger sample of red giant stars, as well as a compiled list of over 31 million stars with spectroscopic line-of-sight velocity measurements. 
  While much less evident in these two older samples, the distribution of stars in action-configuration space suggests that resonance features are present here as well. The position of the feature in action-configuration space suggests that the new feature may be related to the Galactic bar, but other possibilities are discussed.
  }

    \keywords{Galaxy: kinematics and dynamics -- Galaxy: structure -- Galaxy: disc -- Stars: variables: Cepheids}
 
   \maketitle
%

\section{Introduction}


The recent third public release of the \gaia{} survey \citep[][hereafter  \gaia\ DR3]{GaiaCollaboration:2016,gdr3_vallenari}, has provided the community with the largest ever homogeneous set of spectroscopic line-of-sight velocities (\vlos{}) for nearly 33 million stars in the Milky Way, supplementing the high precision astrometry provided already in the \gaia\ Early Data Release 3 \citep[][hereafter \gaia\ EDR3]{GaiaCollaboration:2021release}. Using the new astrophysical parameters also provided in \gdrthree, \citet[][hereafter \drimmelgaia{}]{Drimmel22_gaia} selected a large sample of red giant branch (RGB) stars to map the kinematics of the Galactic disc over an unprecedented volume of the Galactic disc. The stand-out feature of the these velocity maps was the clear signature of a quadrupole pattern in the Galactocentric \vrad{} component in the inner disc (their Figure 16). This bisymmetric feature in \vrad{} is just the expected signature of a galactic bar with mean inward and outward motion on either side of its major axis. Even more remarkable were clear signatures of large scale non-axisymmetry in \vrad{} visible all the way out to the outer disc, at Galactocentric radius $R > 10 $ kpc. For the first time, we can clearly compare the inner and the outer Galaxy on the same velocity maps, and show the influence of the bar out to at least the outer Lindblad resonance (OLR), found to be beyond the solar circle. Specifically, \drimmelgaia{} infer the Milky Way's bar to have a pattern speed of $\Omega_{\rm bar} = 38.1$ \kmskpc, with a corotation radius $R_{\rm CR} = 5.4$ kpc. This outcome is consistent with previous studies based on earlier \gaia\ data releases \citep[][]{PerezVillegas2017, Monari:2019}, confirming that the Milky Way has a long rather than a short bar, as previously thought, where the outer Lindblad resonance (OLR) was thought to coincide with the solar neighbourhood \citep{Dehnen2000resonance}.

Of course, it has long been known that the Galactic disc is not axisymmetric. Large scale spectroscopic and photometric surveys, even with sparse spatial coverage, already hinted at streaming or bulk motion in the disc, with trends in both Galactic height and radius \citep{Widrow:2012,Carlin:2013,Williams:2013,bovy2015_psd,khanna1,dr2kinmap2018}.
The first opportunity to study this in a homogeneous manner presented itself soon after the \gaia'a second data release (\gdrtwo{}), using which \cite{Antoja2018} discovered the presence of large-scale diagonal ridges in the \vtan{}-R density space, which are even more striking when mapped by the Galactocentric velocities, \vrad{} and \vz{} \citep{Ramos:2018,Fragkoudi:2019,Khanna:2019,Laporte:2019,Bernet:2022,Lucchini:2022}. Additionally, \cite{Antoja2018} also discovered an overdensity, known commonly as the phase-spiral, in the $z$-\vz{} plane, when mapped by \vrad{} or \vz{}. This spiral pattern, thought to be a result of a perturbation to the Galactic disc, has since been dissected in the chemo-dynamic space,  \citep{Bland-Hawthorn:2019,Xu:2020,Li:2021} in order to study its origins. With \gdrthree{}, this feature is now mappable over a much larger extent of the disc \citep{Recio-Blanco22_gdr3,Hunt:2022}.

\citet{trick:2018,Trick:2019} and \citet[][hereafter T22]{Trick2022angle}, further explored the \gdrtwo{} dataset in action-angle space. Actions are invariant quantities for a steady or slowly varying axisymmetric Galactic potential. Along with the total energy ($E$), these form a set of invariant quantities, that can be used to trace phase-mixed substructure in the Galaxy \citep{Kalnajs:1991,Ting:2019,Monari:2019,Malhan:2022}. The radial action (\jr{}) is a measure of the eccentricity of a star's orbit, the vertical action (\jz{}) measures the excursion away from the Galactic plane, while \jphi{} is the z-component of the angular momentum, which can be used to infer the guiding radius of a star's orbit. In particular, \citet[][hereafter T21]{Trick:2019} plotted the distribution of \jr{} against \lz{}, to show the presence of large diagonal overdensities in this space. The features seen by T21 are manifestation of the ridges discovered by \cite{Antoja2018}. By considering the action-angle space, these features could be directly linked to a series of expected resonances of the Galactic bar. The continued efforts of several independent studies suggest that some features may not be entirely due to the action of the bar, but may also include spiral arm resonances or perturbations from satellite passages in the recent Milky Way history \citep{Hunt:2018,Hunt:2019,Quillen2018,Fragkoudi:2019,Trick:2019,Martinez-Medina:2019,Khanna:2019,Khoperskov:2021,Antoja:2022}.

The availability of 6D phase-space data for 33 million stars not only allows us to map the kinematics over a huge volume, but also to study the differences between features traced by various populations. In this contribution, we use the sample of the young Cepheids presented in \drimmelgaia{} to probe the disc kinematics to large distances and galactocentric radii.  While being a relatively small sample, they have the advantage of having excellent relative distance errors that allow us to also accurately determine their individual velocities to large distances. In this sample we note a resonance-like feature in their kinematics in the outer disc. We then turn to much larger samples of older stars with complete phase-space information to see if they too have evidence of a resonance feature in the outer disc.

The paper is organised as follows. In Sect. \ref{sec:data_methods} we briefly describe the three datasets used, and our analysis methods. In Sect. \ref{sec:outer_res} we present evidence for a new resonance-like feature in the outer disc, then discuss further its relation to known resonance features in Sect. \ref{disc}. We summarise our findings in Sect. \ref{sum}.

\section{Data and Methods}
\label{sec:data_methods}

\begin{table}
\centering
\caption{The number of stars contributed by the individual surveys to our extended \rvs{} sample. All stars here satisfy the distance uncertainty condition in \autoref{eqn:cbj_uncertainty}. \label{tab:vlos_numb}}
\begin{tabular}{l|l|l}
\hline
Survey  & $N$ & $N \quad(|z|<1$\kpc{}$)$   \\
\hline
\textit{\gdrthree{}}  & 32,390,397 & 30,142,671\\
\textit{\lamost{} $LR$}  & 1,876,313 & 1,295,921 \\
\textit{\lamost{} $MR$}  & 29,168 & 24,977 \\
\textit{\apogee{}}  & 85,994 & 72,086\\
\textit{\galah{}}  & 17,673 & 15,198\\
\textit{\rave{}}  & 7556 & 7239\\
\hline
Total & 34,407,101 & 31,558,092 \\
\hline
\end{tabular}
\end{table}

\subsection{Dataset}
We restrict our analysis to three data subsets constructed primarily using \gdrthree{}. We select both young and old population of stars to illustrate the similarities and differences in their kinematics. 

\textbf{Cepheids:} To trace the young population we adopt the sample of classical Cepheids published by \drimmelgaia{}. This consists of 1948 stars with estimated ages younger than 200 Myr and that have spectroscopic line-of-sight velocities, and their distances are estimated based on the period-Wesenheit-metallicity (\textit{PWZ}) relation \citep{Ripepi:2019,Ripepi:2022}. The relative distance error of this sample is less than 6.25\% for 90\% of this sample. Thanks to the precision in the distances, the uncertainties in the velocities perpendicular to the line of sight, derived from the proper motions and distances, are less than 10\kms{} for 90\% of our sample to a distance of 6\kpc, while the median uncertainty at this distance is less than 5\kms. Meanwhile, the median uncertainty in the spectroscopic line-of-sight velocities remain well below 5\kms at all distances. The resulting uncertainties in the galactocentric azimuthal velocity is less than 10\kms for 90\% of our sample to a distance of at least 9\kpc from the Sun.

\textbf{Red Giant Branch (RGB):} To trace the older population, similarly, we adopt the exact same sample of nearly 5 million red giants used by \drimmelgaia{}. The full details of the sample selection can be found in their paper, but is essentially based on the position in the Kiel diagram of sources that are provided with stellar atmospheric parameters in \gdrthree{}. The distances for this sample are taken from the `photogeo' distance catalogue of \cite{Bailer-Jones:2021}, and as in \drimmelgaia{} we restrict our sample to within 1\kpc of the Galactic plane.

\textbf{Radial Velocity Sample (RVS):} Finally, in order to construct the largest possible sample with full velocity information, we follow the scheme laid out in \cite{Khanna:2022}. We begin by selecting all sources in \gdrthree{} with a valid {\tt radial\_velocity}. The recommended correction to the {\tt radial\_velocity\_error} for these stars was then applied following \cite{2022arXiv220605989B}. We supplement this sample with publicly available \vlos{} measurements from the following spectroscopic surveys; \lamost{} DR7 Low Resolution \citep[LR,][]{lamost}, and Medium resolution \citep[MR,][]{Lamostmrs} surveys, \rave{} DR6 \citep{Steinmetz:2020}, \galah{} DR3 \citep{Buder:2021}, and \apogee{} DR17 \citep{abdurrouf2022a}. To the \lamost{} LR sample, we apply a $+7.76$ \kms{} offset as recommended by their release note\footnote{https://dr7.lamost.org/v2.0/doc/release-note}. For those stars where a \gdrthree{} \vlos{} is unavailable, we assign velocities in the following order: \galah{}, \apogee{}, \rave{}, and \lamost{}. This is in accordance with the typical \vlos{} uncertainty in these surveys. We again use the `photogeo' distances from \cite{Bailer-Jones:2021}, for this sample, but restrict only to high quality distances, requiring,
\begin{equation}
\label{eqn:cbj_uncertainty}
 \frac{0.5 \times (r\_hi\_photogeo - r\_lo\_photogeo)}{r\_med\_photogeo} < 0.2 .
\end{equation} With this cut we remove about 1.5 million stars from the entire sample. While the selection process for the \rgb{} sample was more sophisticated, applying a distance cut as above is one way in which we can restrict to an \rvs{} sample with trustworthy distances. Lastly, just as for our \rgb{} sample, we restrict to within 1\kpc of the Galactic plane, resulting in a total of  31,558,092 stars in our final \rvs{} sample.  In \autoref{tab:vlos_numb}, we list the contribution from individual surveys that satisfy the distance quality as well as the |z| selection.

\subsection{Action estimation}
In general, actions can be computed analytically only for potentials where the variables can be separated, in order to solve the Hamilton-Jacobi equation \citep{Sanders:2016}. Spherical \& St\"{a}ckel potentials \citep{deZeeuw:1985}, are examples where this is possible. For more general axisymmetric potentials, \cite{Binney2012stackel} developed a method where the Galaxy's gravitational potential is assumed to be similar to a St\"{a}ckel potential. This, so called \textit{St\"{a}ckel-fudge} method is implemented in the \textit{AGAMA} code \citep{agamapaper}, to calculate the classical axisymmetric actions $J=$(\jr{},\jphi{},\jz{}) mentioned in the Introduction.
The actions estimated this way remain conserved over time and serve as  \textit{`true'} integrals of motion only in an axisymmetric potential. For a more realistic galactic potential (e.g., the Milky-Way), with clear non-axisymmetric features such as a bar and spiral arms, this is not the case. However, combining the \textit{St\"{a}ckel-fudge} method with an axisymmetric potential well fitted to observed data still allows us to estimate the instantaneous axisymmetric actions, and extract the main features and substructure of the underlying kinematics in the Galaxy. 
Essentially, any substructure that appears in this action space is likely due to non-axisymmetric or time varying components of the gravitational potential, whether or not the actions are accurate or not. 
 We use the \textit{AGAMA} code to compute the instantaneous set of actions, $J=$ (\jr{},\jphi{},\jz{}), in three different axisymmetric potentials: a) \cautunpot{}, b) \mcmillanpot{}, and c) \bovypot{} \citep[][respectively]{cautun20,McMillanpot2017,Bovy:2015}. 

\subsection{Coordinate transformations}
We follow the scheme used in \drimmelgaia{} to transform sky coordinates, proper motions and distances to Galactocentric positions and velocities. That is, to compute the azimuthal components of the Sun's Galactocentric velocity, we use the precise measurement by \cite{Reid2020}, of the proper motion of Sgr A$^{*}$, i.e., $(\mu_l, \mu_b) = (-6.411 \pm 0.008, -0.219 \pm 0.007)$ \masyr. Meanwhile, the ESO Gravity project's most recent measurement of the orbit of the star S2 around the the Milky Way's supermassive black hole yields both a very precise distance to the Galactic centre, $R_\odot = 8277 \pm 9$(stat) $\pm 30$(sys) pc, as well as the line-of-sight velocity towards Sgr $A^{*}$ \citep{GravityCollaboration:2021}. This gives, for the solar velocity with respect to the Galactic centre,
\begin{equation}
{\bf{v}}_\odot =
\begin{pmatrix}
9.3 \pm 1.3  \\
251.5 \pm 1.0  \\
8.59 \pm 0.28 
\end{pmatrix} \kms
\end{equation} assuming that Sgr A$^{*}$ is stationary with respect to the Galactic centre. The cylindrical coordinate angle $\phi={\rm tan}^{-1}(Y/X)$ increases in the anti-clockwise direction, while the rotation of the Galaxy is clockwise. The height of the Sun above the Galactic plane is assumed to be 0. The heliocentric Cartesian frame is related to Galactocentric by $X_{\rm hc}=X+R_{\odot}$, $Y_{\rm hc}=Y$ and $Z_{\rm hc}=Z$. $X_{\rm hc}$ is negative toward $\ell=180^\circ$ and $Y_{\rm hc}$ is positive towards Galactic rotation.

\section{Signature of an outer resonance}
\label{sec:outer_res}

\subsection{Kinematics of the Cepheids}
\label{sec:kin_Ceph}

\begin{figure*}
    \centering
  \includegraphics[width=1\textwidth]{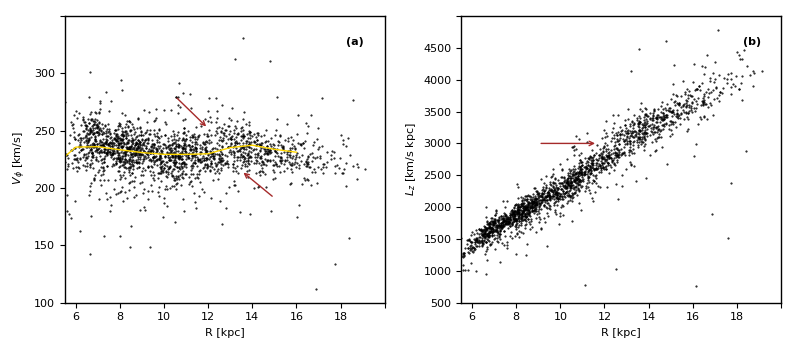}  
    \caption{The observed azimuthal velocities (panel a) and specific angular momentum (panel b) of the young Cepheids. The yellow curve in panel (a) traces the median rotation curve for the sample. The arrows in both panels indicate the apparent gap in the distribution and the location of a new resonance-like feature, around \lz{}$\sim3000$ \lzunit.
    \label{fig:azVel_and_Lz_vs_R}}
\end{figure*}

\begin{figure}
\includegraphics[width=1.\columnwidth]{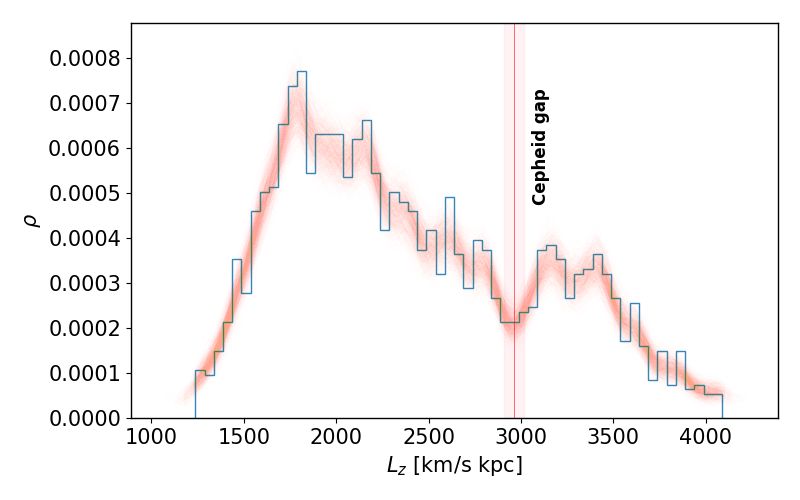} 
\caption{Distribution of the z-component of Angular momentum (\lz{}) for the Cepheid sample, shown as blue solid line. In the background in orange, is overplotted the KDE estimates of 1000 bootstrapped samples. The solid orange vertical line (and the shaded region), marks the position of the dip in the \lz{} distribution at $2950\pm46$\lzunit. This corresponds to the gap marked in \autoref{fig:azVel_and_Lz_vs_R}. \label{fig:cepheid_angmom_kde}}
\end{figure}


We here consider the kinematics of the young Cepheids from \drimmelgaia{}. Plotting their azimuthal velocity with respect to Galactocentric radius (i.e., the rotation curve), we note a gap in the azimuthal velocity distribution at approximately $R=13$\,kpc (See Fig. \ref{fig:azVel_and_Lz_vs_R}(a)). This gap has a negative slope with respect to $R$, between about 12\kpc{} to 14\kpc. However, plotting the angular momentum of the Cepheids (Fig. \ref{fig:azVel_and_Lz_vs_R}(b)) shows that this gap is apparently at a fixed value of \lz{($=V_\phi R$)}. 
In \autoref{fig:cepheid_angmom_kde} we show the 1D distribution in \lz{}, for the Cepheid sample, where the gap is quite clearly identifiable in this space. To estimate the precise location and its width, we perform a kernel density estimate (KDE) using the \textsc{sklearn} package \citep{scikit-learn}. We use a \textit{Gaussian} kernel, setting the bandwidth following \cite{Scott:1992}, and perform the KDE on 1000 bootstrapped samples. In \autoref{fig:cepheid_angmom_kde}, this is shown as the set of smoothed histograms in the background (orange). For each of the 1000 samples, we estimate the minimum between $2800 < L_{z} < 3100$. We find that the dip in the Cepheid \lz{} distribution occurs at \lz{} $=2950\pm46$ \lzunit. Because this gap is well defined and at a fixed value of angular momentum, a conserved quantity, it has the characteristic of being a resonance feature. 
However, as we will discuss further in Sect. \ref{sum}, this interpretation is problematic.

We also note that this gap coincides with an apparent "bump" in the rotation curve of the Cepheids derived from radially binning their azimuthal velocities (See \ref{fig:azVel_and_Lz_vs_R}(a)). However, this bump is just an artefact of the noted gap itself. We therefore locate the radius of this resonance-like feature, using instead the $L_Z$ distribution, which is very well described by a simple linear fit. We find $L_Z = 231.4 R + 9.65$\lzunit, with a correlation of 0.87. 
We note that the intercept is insignificantly small.  Dropping this last term, we can therefore adopt $R_g \equiv L_Z/231.4$\kpc as the guiding radius for each star.  The location of this feature is then taken as the guiding radius for $L_Z = 2950$\lzunit, that is at $R_g=12.75$\kpc. In section \ref{disc} below, we discuss the location of this feature and its relation with the known resonances of the bar. 

As \lz{} correlates well with $R$, one might wonder whether a gap in the radial distribution might cause the feature seen at \lz{}$=2950$\lzunit.  The feature is indeed found in a restricted range of radii, as the \lz{} distribution (Fig. \ref{fig:azVel_and_Lz_vs_R}a) has a finite width in radius. However, in that same radial range we see Cepheids with both lower and higher values of \lz{}. The distribution of \lz{} is nevertheless modulated to some degree by the spatial distribution of Cepheids in the Galactic plane. In fact, \citet{Poggio:2021} and \drimmelgaia\ point out that the Cepheids in the outer disc trace nicely an outer spiral that is seen in HI, as modelled by \citet[][]{Levine06}. However, this arm is mainly in the third quadrant, yet we also see the gap in the \lz{} distribution for the Cepheids in the first and second quadrants. 

We have also checked possible selection effects that might cause this gap. We can exclude extinction effects, as Cepheids with a given \lz\ are not restricted to a small area on the sky, but span a large range of Galactic azimuth. Also, extinction would indiscriminately remove Cepheids for a large range of \lz. Comparing to independent high-fidelity Cepheid catalogues in the literature, it is estimated that our \gdrthree\ sample of Cepheids is of the order of being 90\% complete \citep{RipepiDR32022}. This likely varies depending on the number of epochs available for characterising the variability, which is determined by \gaia's scanning law. However, this will only introduce an on-sky directional dependence which, like extinction, will not selectively remove Cepheids at a particular distance or angular momentum.

%
\subsection{Kinematics of the RGB \& RVS samples}

\begin{figure*}
\includegraphics[width=1\textwidth]{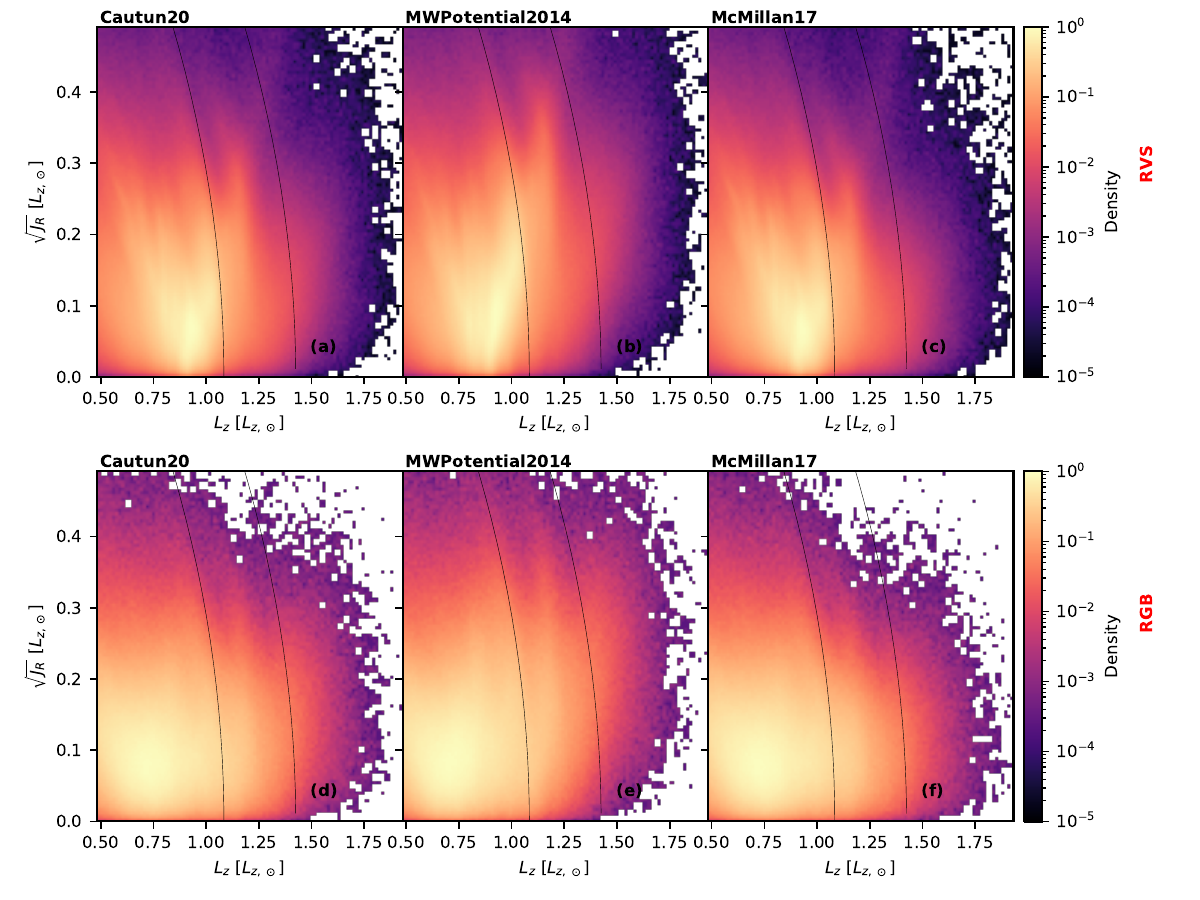}
\caption{Distribution in ($\sqrt{J_{R}}$,\lz{}) plane for the \rvs{} (upper panels) and the \rgb{} (lower panels) samples, computed using \agama{} for four different gravitational potentials. Both datasets show diagonal ridges, but the \rgb{} sample extends out further into the outer disc. Beyond $1.42$ \lsun{} the density falls off sharply which makes it harder to pick out features.The black curves overplotted mark the expected OLR of the bar at $1.08$ \lsun{}, and the Cepheid gap at $1.42$ \lsun{}.}
\label{fig:actions_rvs_rgb_dens}
\end{figure*}

\begin{figure*}
    \centering
   \includegraphics[width=1\textwidth]{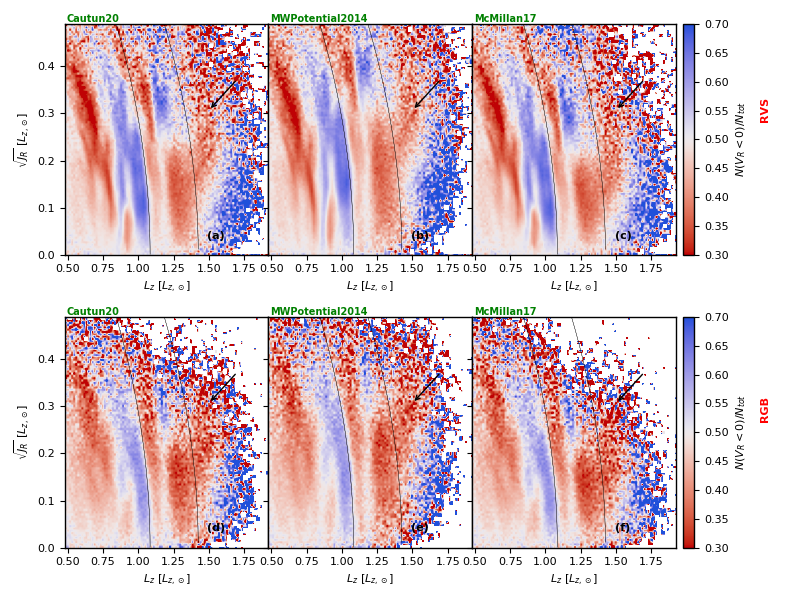}
    \caption{Distribution in the $(\sqrt{J_{R}}$,\lz{}) plane for the \rvs{} (upper panels) and the \rgb{} (lower panels) samples, computed using \agama{} for four different potentials. Here we colour code the distribution by the fraction of stars moving inward (\vrad{}$<0$). The black curves overplotted mark the expected OLR of the bar at $1.08$ \lsun{}, and the Cepheid gap at $1.42$ \lsun{}. The diagonal ridges are much clearer in this space compared to \autoref{fig:actions_rvs_rgb_dens}. We report the presence of a previously unseen new overdensity of outward moving stars just beyond the Cepheid gap, marked by black arrows.} \label{fig:actions_rvs_rgb_nvr}
\end{figure*}

The Cepheid sample have revealed an interesting feature in the \lz{} distribution in \autoref{fig:cepheid_angmom_kde}, suggesting the presence of a resonance. 
We now turn to the \rvs{} \& \rgb{} datasets to gain a many-fold increase in density (from 1000s to millions of stars), allowing us to explore the distribution in action space where we expect to see more clearly the signature of resonances. In \autoref{fig:actions_rvs_rgb_dens}, we plot the distribution of the two datasets in the ($\sqrt{J_{R}}$, \lz{}) plane for the four different axisymmetric potentials listed in \autoref{sec:data_methods}. We normalise both axes by the z-component of angular momentum at the Sun, \lsun{}$=2081.6$ \lzunit. We recover the large scale diagonal features (ridges) seen in this space by T21. (Unlike T21 we choose to present our results using $\sqrt{J_{R}}$ and not ${J_{R}}$ itself, as it enhances the features slightly.) The \rvs{} sample, being dominated by nearby bright stars, shows a high concentration near \lsun{}. The \rgb{} distribution is comparatively diffuse, while retaining most of the ridge features. We also note that while there are subtle differences between the individual potentials, overall these map similar features. In this regard, the \bovypot{} seems to be least consistent with the other two potentials. We think this is due to the difference in the circular velocity normalisation between the potentials. In particular, a group of stars that rotate close to the circular velocity of a chosen potential, will be on near-circular orbits and thus have lower \jr{}. Since, the circular velocity in the \bovypot{} is about 220 \kms{}, while in the other two it is around 230 \kms{}, this would explain the vertical shift to lower \jr{} in \bovypot{}.

Compared to T21, we have the benefit of adding in more data in the outer disc thanks to \gdrthree{}. Keeping \autoref{fig:cepheid_angmom_kde} in mind, we are interested in probing the region around \lz{}$=2950$ \lzunit\ (or $\sim1.42$\lsun). However, it is clear from \autoref{fig:actions_rvs_rgb_dens}, that even with the added coverage, the density in action space falls off sharply beyond this \lz{} value. 
Nevertheless, no clear gap in the distribution is seen. Thus, we again follow T21, and in \autoref{fig:actions_rvs_rgb_nvr}, show the same distribution in the ($\sqrt{J_{R}}$,\lz{}) plane, but now mapped by $N(V_{R}<0)/N_{tot}$, i.e., the fraction of sources moving in towards the Galactic centre. This has two immediate effects; first, it makes the diagonal features stand out dramatically, and second, we are now able to observe features for the entire extent of our datasets. Our \autoref{fig:actions_rvs_rgb_nvr} can be directly compared to Figures 1, 11 \& 12 in T21, who were limited in coverage out to about \lz{}$/$\lsun{}$=1.4$. For their analysis T21 used the \textit{galpy} \citep{Bovy:2015} code, using the \bovypot{} to compute the actions, i,e, comparable to \autoref{fig:actions_rvs_rgb_nvr}(c). T21 and T22 also showed that the ridges in action space can be roughly traced with lines of negative slope of about $-1$ in ($J_{R}$,\lz{}), at different locations in \lz{}, corresponding to the various resonances. The purpose of including this slope is only to serve as a guide to the reader.

In \autoref{fig:actions_rvs_rgb_dens} we plot two such lines (curves in $\sqrt{J_{R}}$,\lz{}). The first one is at the expected location of the bar's OLR, based on \drimmelgaia{}, around $R_{g}=9.7$ kpc, or at 1.08 \lsun{} ($231.4 \times R_{g}/$ \lsun{}). As \autoref{fig:OmegaRGB} shows, however, the location of the OLR has an uncertainty associated with it (shaded region), and could be as large as $R_{g}=10.2$ . Here, we choose the median value of $R_{g}=9.7$ kpc, but choosing a higher value, as well as a different choice of the circular velocity at the Sun, will shift the expected location of the resonance.  The second curve is at the Cepheid gap presented in \autoref{fig:cepheid_angmom_kde} at $1.42$ \lsun{}. The OLR ridge indeed seems to be traced well with the first curve, though we note that the ridge has a width which is also expected \citep[see for ex:][]{Binney:2020}. The second curve corresponding to the new feature is harder to link to any ridge. Similarly, we also overplot the two curves on \autoref{fig:actions_rvs_rgb_nvr} which maps $N(V_{R}<0)/N_{tot}$. In each potential the OLR curve seems to mark a boundary between inward and outward systematic motions. This is consistent between both the \rvs{} and the \rgb{} samples (though as before, more diffuse in the latter). 

The classically expected orbital behaviour around the OLR \citep{Weinberg1994, Dehnen2000resonance,sellwood2010resonance}, was illustrated in action space by the test particle simulations of T21 \& T22. In particular, 
since the Galactic bar is leading the Sun by about 20 degrees, 
the stars inside the OLR curve (lower \lz{}) are expected to move outwards (red), while those outside the OLR curve (higher \lz{}) would be expected to move inwards (blue). However, the observed behaviour is opposite to this expectation. This was also remarked upon by T21 \& T22. In particular, in their analysis using angle space (instead of actions) they showed that a pattern speed close to our adopted value from \drimmelgaia{} seems to be the most favourable candidate to explain the expected velocity distribution around the OLR, though it was not clear why the orientation was flipped in action space. 
Notwithstanding this open question regarding the `correct' orientation of the red-blue feature at resonances, given that our predicted location for the OLR marks a boundary in this space, we will continue to refer to it as the OLR. 

The location of the second curve is quite interesting for two reasons. First, inwards of the second curve, is the dataset that was mapped by T21. In particular, they were the first to show the ridge just inside of this curve. We are now able to present a new feature of net positive radial motions just outside this second curve (marked by black arrows). That is, we find a clump of outward moving stars, with high radial action ($\sqrt{J_{R}}>0.15$ in \mcmillanpot{}), though as before we note subtle differences between the individual potentials. 


Our coverage extends out to almost \lz{}$/$\lsun{}$=2$. In the region beyond \lz{}$/$\lsun{}$>1.7$, there is a notable feature of negative radial motions (in blue) in \autoref{fig:actions_rvs_rgb_nvr}. This could be another interesting feature worth exploring, as we see it regardless of which distance estimator is used, those from \cite{Bailer-Jones:2021}, or inverse parallax (not shown here). We do however, remind the reader of possible artefacts at the edge of our data coverage (see also \drimmelgaia{}).

%
\section{Discussion}
\label{disc}

\subsection{Possible 1:1 resonance with the bar}

\begin{figure}
    \centering
    \includegraphics[width=0.49\textwidth]{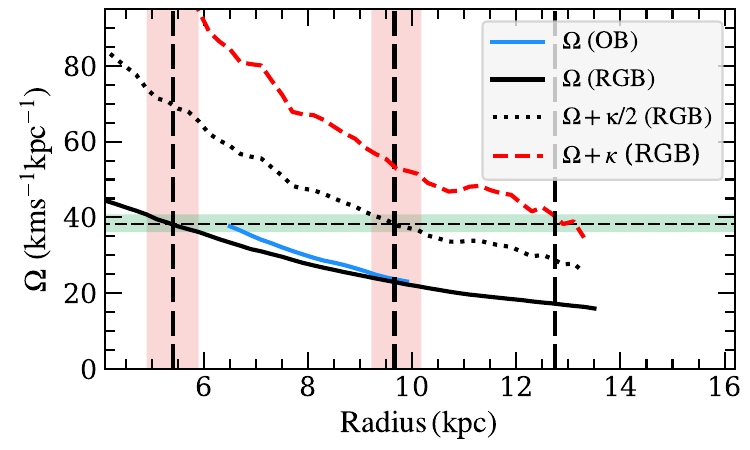}
    \caption{The angular velocity of the RGB (solid black curve) and OB stars (solid blue curve). The dashed vertical lines mark the estimated position of corotation (5.4\kpc) and the Outer Lindblad Resonance (9.7\kpc), as estimated in \drimmelgaia{}, and the new resonance-like feature at 12.75\kpc. The thin horizontal dashed line indicates the pattern speed inferred from the corotation radius. We note that the radius of the new resonance-like feature is quite close to the radius where $(\Omega + \kappa)$ is equal to the pattern speed of the bar, that is, approximately at the expected radius of the bar's 1:1 resonance.}
    \label{fig:OmegaRGB}
\end{figure}

As mentioned above, given that we observe a gap in the azimuthal velocity radial distribution that is at a fixed value of \lz{}, a conserved quantity for an axisymmetric potential, we believe this feature might be due to a resonance. 
We now discuss the position of this feature with respect to the other known resonances in the disc. Figure \ref{fig:OmegaRGB} shows the angular velocity of the RGB and OB stars, and the corotation and the OLR of the bar, according to \drimmelgaia{}. We note that the position of the resonance feature at $R_g=12.7$\kpc, is near the expected position of the bar's 1:1 resonance.

\begin{figure}
    \centering
    \includegraphics[width=0.49\textwidth]{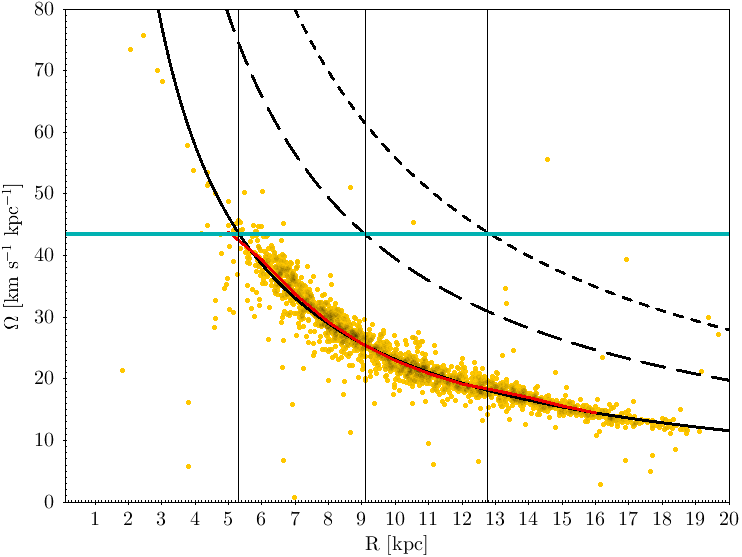}
    \caption{The angular velocity of the Cepheids, using $\Omega(R) = L_Z(R=R_g)/R^2 = 231.4/R$ (solid black curve), with the angular velocity of the individual Cepheids as yellow points. The overplotted red curve is the angular velocity that would result from using the Cepheid rotation curve from \drimmelgaia{}. Dashed curves are $\Omega + \kappa/2$ (long dash) and $\Omega + \kappa$ (short dash).  The right-most vertical line is at $R=12.75$\kpc, the position of the new resonance-like feature, and the horizontal turquoise line is the resulting estimated pattern speed of the bar,     assuming this feature to be the 1:1 bar resonance. The other vertical lines mark the resulting positions for corotation and the OLR. 
    }
    \label{fig:OmegaR}
\end{figure}

The clarity and sharpness of this outer feature in \lz{}-$R$ space motivate us to hypothesise that this is a 1:1 resonance feature, and to
derive the pattern speed of the bar accordingly. We use the angular velocity of the Cepheids rather than that of the RGB stars as done in \drimmelgaia{}, as it will be closer to the actual circular velocity of the disc thanks to the youth and low velocity dispersion of this sample. In Fig. \ref{fig:OmegaR} we show the resulting $\Omega(R_g)$ curve for the Cepheids. 
From \citet{BT2008} (eq 3-59), in the epicyclic approximation the epicyclic frequency is:
\begin{equation}
    \kappa^2(R) = \left( R \frac{d \Omega^2}{dR} + 4 \Omega^2 \right)_{R_g}\, .
\end{equation}
Taking $\Omega^2 = V_\phi^2/R^2 = L^2_Z/R^4$ and $L_Z(R_g) = 231.4 R_g$ (see section \ref{sec:kin_Ceph}), we find $\Omega(R_g) = 231.4/R_g$ for the angular velocity of the Cepheids and the epicyclic frequency is then $\kappa(R_g) = \sqrt{2} \cdot 231.4/R_g$.  Assuming the 1:1 resonance is at $R_g=12.75$\kpc results in a bar pattern speed of 43.5\kmskpc, with the 2:1 OLR and corotation at 9.1\kpc and 5.3\kpc, respectively. This corotation radius in good agreement with that of \drimmelgaia{}, while the radius of the OLR is about a half kiloparsec away from that estimated using the RGBs. 

\begin{figure}
    \centering
    \includegraphics[width=0.49\textwidth]{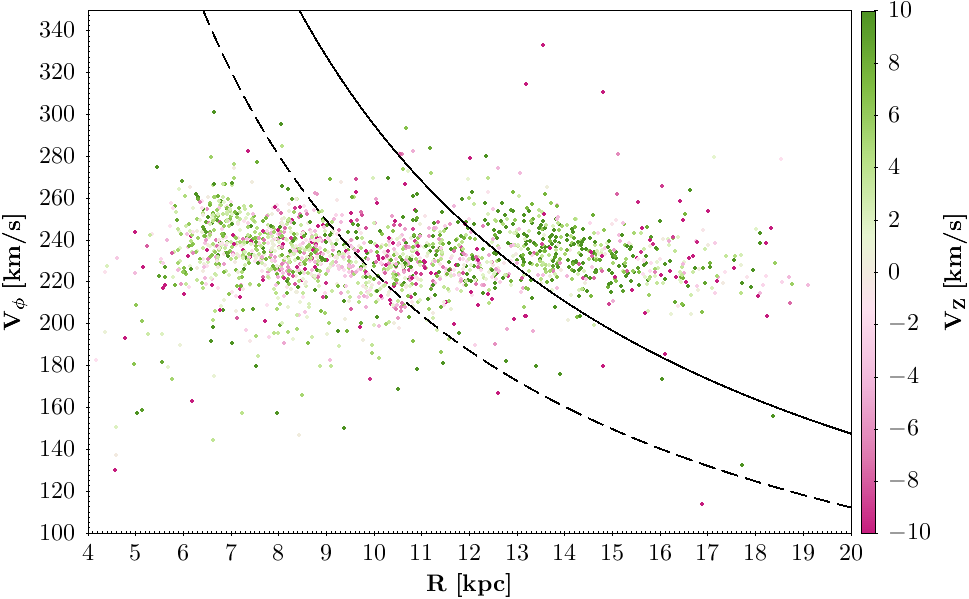}
    \includegraphics[width=0.49\textwidth]{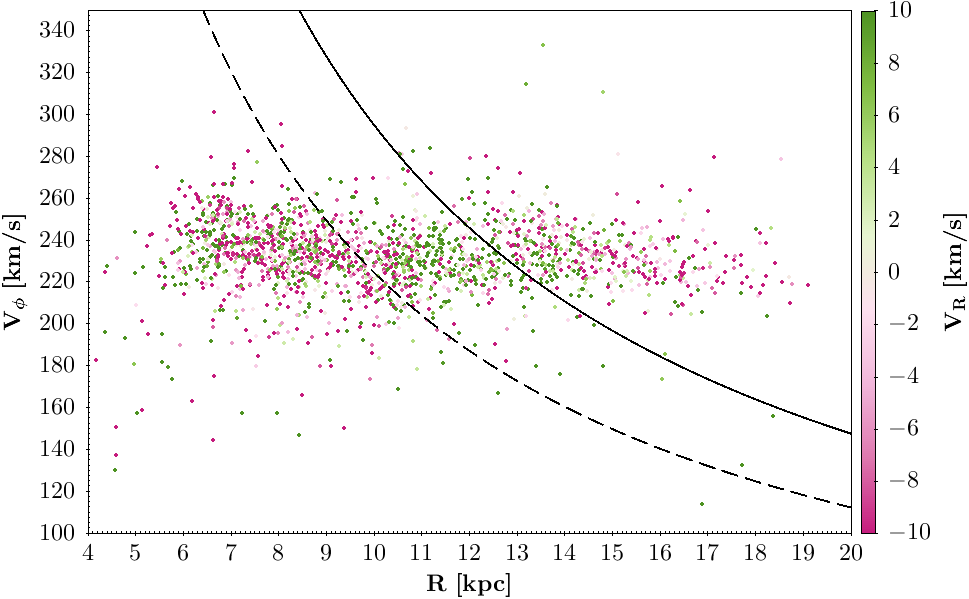}
    \caption{The observed azimuthal velocities of the young Cepheids, with the points coloured with respect to their galactocentric vertical (upper plot) and  radial velocities. Solid and dashed curves mark lines of constant $L_Z$ of the 1:1 and 2:1 OLR, respectively.
    }
    \label{fig:VzVr_vs_R}
\end{figure}

Up to this point we have only discussed the kinematics of the Cepheids associated with their azimuthal velocities. However, it is worth noting what is seen in their other two velocity components with respect to galactocentric radius and this new resonance feature. Figure \ref{fig:VzVr_vs_R} shows again $V_\phi$ with respect to $R$, but with the $V_Z$ and $V_R$ velocities indicated in colour in the upper and lower panels. Also shown are the constant $L_Z$ curves for the 1:1 and 2:1 (OLR) resonances, taking $L_Z$ of the OLR to be at $R_g=9.7\kpc$, that is $L_Z= 2244.6 \lzunit$. Since these plots integrate over a large range in $\phi$ we might not expect any clear pattern, however in the outer disc we see that the $L_{Z} = 2950$\lzunit\ of the 1:1 resonance marks a clear boundary for a change in \vrad{} \emph{and} $V_Z$.   
That we also see systematic positive $V_Z$ velocities in the part of the outer disc that we are sampling is to be expected: This is just the warp signature that has already been noted \citep[][]{Poggio2018}. What is not so expected is that the $L_Z = 2950$\lzunit\ boundary would so clearly mark the onset of these vertical motions.  

\begin{figure*}
    \centering
    \includegraphics[width=1\textwidth]{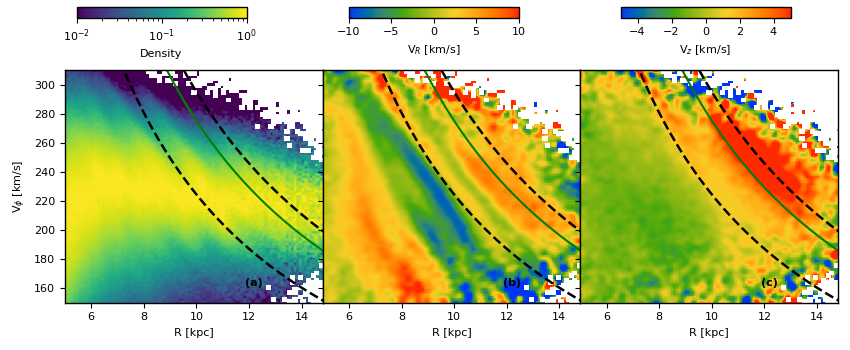} 
    \caption{The \rgb{} sample shown in the \vtan{}-$R$ space in density (panel a), median \vrad{} (panel b), and median \vz{} (panel c). The two black dotted lines are the locations of the expected OLR \& the 1:1 resonance of the Galactic bar. The green solid line marks the location of the discontinuity seen in \acantoja{} and \paul{}. \label{fig:ridges_comb}}
\end{figure*}

For comparison, \autoref{fig:ridges_comb} shows the \rgb{} sample in the \vtan{}-$R$ space, colour-coded by density, the median \vrad{}, and the median \vz{} velocities. The same plots for the full RVS sample (not shown) are very similar.
The much larger number of stars in this sample again allows us to more clearly identify features, for instance those correlated with the OLR. With respect to $V_Z$ we see a similar pattern of systematic positive (upward) velocities in the outer disc, but with a clear difference with respect to the Cepheids in that the $L_Z = 2950$\lzunit\ boundary does not indicate where these vertical motions begin. 
In any case, that curves of constant $L_Z$ in the $V_\phi$-$R$ plane mark the boundary of the onset of systematic vertical motions suggest that the in-plane (epicyclic) and vertical motions are coupled in the outer disc, similar to what would be expected from the perturbation from a passing satellite. Alternatively, the vertical motion may be simply determined by the star's guiding radius, for which \lz\ serves as a proxy.

\subsection{Comparison to previous studies}

\begin{figure*}
    \centering
    \includegraphics[width=1\textwidth]{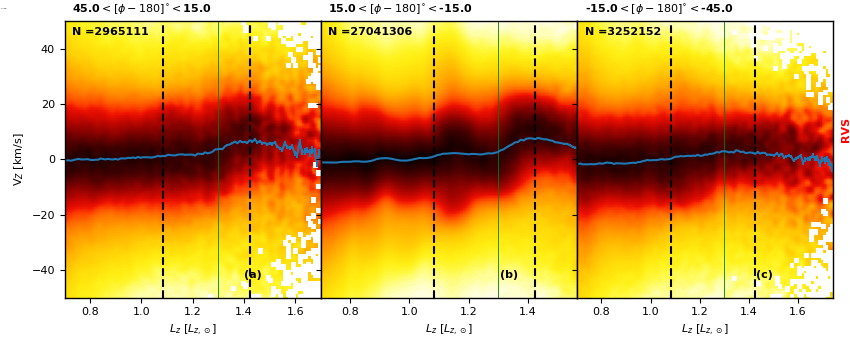}    
    \includegraphics[width=1\textwidth]{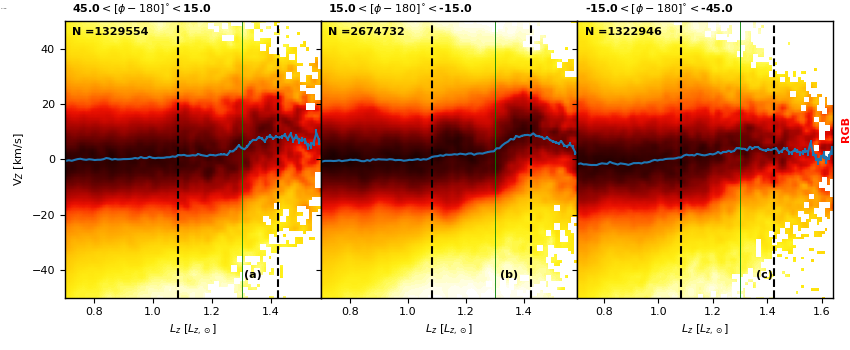}    
    \caption{Distribution of the \rvs{] \& \rgb{} samples in vertical velocity \vz{} against \lz{}. The panels divide the data into three bins of $30^\circ$ width in the azimuth. The middle panel covers stars about 15$^\circ$ either way of the Galactic Anticentre. Two black dotted lines are plotted at the positions of the OLR (1.08 \lsun{}), and at the Cepheid gap (1.424 \lsun{}) in \autoref{fig:cepheid_angmom_kde}. The green vertical line marks the discontinuity (1.3 \lsun{}) noted in recent papers \citep{gaia_anticenter2021,McMillan:2022}}. \label{fig:vz_lz}}
\end{figure*}

The outer disc of the Milky Way is an interesting laboratory for exploring the dynamics and past of the Galaxy. Due to the lower gravitational potential, imprints of perturbations in this region are long-lived, making these still observable today. Over the years, surveys using a variety of kinematic tracers have shown that the outer disc is corrugated and flares with increasing $R$ \citep{Yanny:2013,Xu:2015,Thomas:2019,Bland-Hawthorn:2019,Mackereth:2019}. Using the most recent astrometric data from \gedrthree{}, \citet[][hereafter \acantoja]{gaia_anticenter2021} analysed the kinematics of stars in the Galactic Anticentre region ($170^\circ < l < 190^\circ$). This window is narrow enough, that the line-of-sight velocity can be assumed to be zero, to a good approximation. This allowed \acantoja{} to study the kinematics of a very large sample of stars in the outer disc lacking spectroscopic line-of-sight velocities, and to discover that the velocity distribution is bimodal in this region. More recently, \cite[][hereafter \paul]{McMillan:2022} extended this analysis to a much wider range ($130^\circ < l < 230^\circ$), thus vastly increasing the number density of sources and allowing a study of the variation of this bimodality with respect to Galactocentric azimuth. Together, these works showed that around the 1.3\lsun{}-1.35\lsun{} region, there is a sharp break in the \vz{}-\lz{} space, and that the strength of the feature varies over galactic longitude. 

In \autoref{fig:vz_lz}, we show our \rvs{} \& \rgb{} populations in the $V_{z}-L_{z}$ plane, same as that studied by \acantoja{} \& \paul{}, though here we use the new line-of-sight velocities in \gdrthree. We divide the sample into three bins of $30^{\circ}$ width in the azimuth, tracing either side of the anticentre. All three panels show that with an increase in $R$, the vertical velocity is fairly flat out to $\sim 1.1$\lsun{}. Beyond this guiding radius, the distribution appears bumpy and disturbed as we move further out. The middle panels in \autoref{fig:vz_lz} are centred about the Galactic Anticentre. This coverage overlaps with that of \acantoja{}. It is not surprising then, that we also see a clear break in the velocity distribution around $\sim 1.3$\lsun{}. This feature has a clear dependence with azimuth, again similar to what \paul{} demonstrated. We note that the position of the break seen here is just at a slightly lower \lz{} compared to \acantoja{} \& \paul{}, but this is due to the wider azimuthal bin used here. On each of the panels, we overplot three vertical lines of interest. The two black dotted lines in each panel, are the expected 2:1 OLR ($\sim 1.08$\lsun{}), and the 1:1 ($\sim 1.42$\lsun{}) resonance lines for the pattern speed of the \drimmelgaia{} bar. Additionally, we also overplot in green, the location of the bimodality as observed by us at $\sim 1.3$\lsun{}. The velocity distribution in \autoref{fig:vz_lz} and the overplotted lines of interest, suggest that the Cepheid \lz\ gap we observe in \autoref{fig:cepheid_angmom_kde} is distinct from the bimodality break discovered by \acantoja{} \& \paul{}. Lastly, we also note the presence of a bump-like feature in $V_{z}-L_{z}$, both around the location of the Cepheid gap at $\sim 1.42$\lsun{}, and in the acceptable range for the OLR ($\sim 1.08$\lsun{}$ - 1.17$\lsun{}). We don't draw any conclusions from this, it is likely a mere coincidence, although investigating the signature of vertical oscillations around resonances is interesting in itself.

\section{Conclusions}
\label{sum}

As the quadrupole moment of the bar potential falls off as $R^{-3}$ outside the bar \citep[section 2.4][]{BT2008, Weinberg1994}, it is not expected that the bar would have much influence beyond its OLR. Indeed, we cannot be positively sure that we see a corresponding resonance signature in the RGB and RVS sample, and if we do see one it is very weak. However, we note that \citet{Trick:2019} mentions the possible detection of a weak signature at the 1:1 resonance already in \gdrtwo{} data, based on comparisons with simulations of test particles responding to a barred potential.
Our Cepheid sample however has important distinctions with respect to the older RGB and RVS sample. First, it is dynamically very young at this point in the disc: These stars have not yet completed a single orbit about the Galactic centre. 
Indeed, with ages of less than 200Myr, in the outer disc they should be considered as tracers of the gas from which they were born, as they have not had time to respond to a resonance since their birth. However, also in the gas, we do not expect the 1:1 resonance to manifest itself: Unlike the Lindblad resonances and corotation in the weak bar regime (see sec 3.3 of \citet{BT2008}), we do not expect to see a change in the orientation of closed orbits at this resonance. 
We have also investigated the gas response in the outer disc to a barred potential by looking in more detail at the gas dynamics in already extant simulations of the Milky Way. (See appendix \ref{sec:app}.) Also here, no clear resonance-like feature is manifested at the position of the 1:1 resonance. More appropriate comparisons with simulations of a young stellar population, along the lines of \citet[][]{Pettitt2020}, should be done in the future.

Important characteristics of this sample that helps us to see what would otherwise be subtle features in their kinematics are their excellent distances and their intrinsic low velocity dispersion. The quality of their distances not only assists in accurately assigning a galactocentric radius to each star, but also in deriving an accurate velocity perpendicular to their line-of-sight. 
Meanwhile their low velocity dispersion allows us to see narrow features that would otherwise be "erased" over time in a sample with higher velocity dispersion.  However, if we are seeing a resonance feature, why do we not see a similar feature at the OLR of the bar? 

An alternative explanation is that we are not seeing a resonance feature from the bar but instead a resonance from the spiral arms. For a bar with a pattern speed of $\sim$ 40 \kmskpc, the spiral arms that develop at the edge of the bar have a pattern speed of 30 \kmskpc and their OLR between 11-13 kpc (see Fig.11 of models of a barred Milky Way in \citet[][]{DOnghia2020}. If this new feature is the OLR of the spiral arms, a pattern speed of 31 \kmskpc\ is deduced, which would place the corotation of the spiral arms near the solar circle.
If however the spiral structure that propagates outwards in the disc has a lower pattern speed ($\sim20$ \kmskpc), then our detected feature may be a response to corotation. \citet{Barros2013} argue one should indeed expect a resonance feature at the corotation of the spiral arms, though they report such a feature close to the solar circle, at $R\sim 9$\kpc.
In either case it would indicate that we have identified a spiral arm resonance that overlaps with a bar resonance. 

Recent efforts to deduce a spiral arm pattern speed in the Milky Way have yielded diverse results, possibly complicated by the influence of the bar. \citet{Grosbol2018}, comparing the kinematics of young stars as far as 5 \kpc from the Sun to simulations, find a range of possible pattern speeds between 20 and 30 \kmskpc. More recently \citet{Monteiro2021}, using a sample of young open clusters, deduce a common pattern speed of 28 \kmskpc from four spiral arm segments, supporting the idea of a long-lived spiral pattern. However, \citet{Castro-Ginard:2021}, using the same method and a very similar sample of open clusters, arrive at a completely different conclusion, finding different pattern speeds for multiple arm segments, suggesting that the Milky Way's spiral arms may be transient, as also suggested by the study of \citet{Quillen2018}, who associate kinematic features in a sample of nearby stars to spiral arm crossings. 

Another possibility to be considered is that we are seeing a transient feature from a recent interaction. In their analytic model of a disc-crossing satellite, \citet{Binney:2018} show how the impact of Sgr sets up a large-scale $m=1$ mode in the outer stellar disc. Interestingly, a hole is punched into the disc at the point of transit owing to the in-plane and vertical deflection of local stars. \citet{Bland-Hawthorn:2021} confirm this behaviour for the first time in an N-body simulation (their fig. 7), and show how the hole is sheared into a strong stellar underdensity at or near the impact radius, here assumed to be 20 kpc. The deep gap occurs between the spiral arms generated by the interaction as they wind up slowly after the event, but the depth of the transition is slowly filled in by orbit migration. This gap remains for a few rotation periods and could reasonably account for what is seen here. However, essentially all Sgr orbit studies to date indicate that the disc transit occurred much further out than the observed Cepheid gap \citep[e.g.][]{Laporte:2019}

\gaia\ continues to reveal a surprising amount of structure in the phase space of stars. With each
data release, we are seeing 
increased coverage of the Milky Way's disc which has led to new kinematic features being discovered at each step. The improved quality of the \gaia\ data across the disc allows for better discrimination between the non-axisymmetric features responsible for resonances, and potentially to distinguish these from transient features excited by satellite interactions, which should show significant variation at different azimuthal angles. 
From the \gdrthree\ treasure trove, we have found yet another piece of the Galactic puzzle, one which may help us to identify and disentangle the various dynamical processes shaping the disc of the Milky Way.

\begin{acknowledgements}
RD thanks Ortwin Gerhard for useful discussions. SK acknowledges Tom\'{a}s Ruiz-Lara \& Eduardo Balbinot for preparations of the extended radial velocity catalogue, and Eugene Vasiliev for useful discussion on actions.

RD, VR, GC and TM are supported in part by the Italian Space Agency (ASI) through contract 2018-24-HH.0 and its addendum 2018-24-HH.1-2022 to the National Institute for Astrophysics (INAF).

SK, TCG and ACG acknowledge support from the European Union's Horizon 2020 research and innovation program under grant agreement No 101004110.

RD and SK acknowledge Ricky Smart \& Luciano Nicastro for use of the computing resources under the GLORIA-project, funded by the European Union 7th Framework Programme under grant agreement n. 283783.

PR acknowledges support from the University of Barcelona, via a Margarita Salas grant (NextGenerationEU).

TTG acknowledges partial financial support from the Australian Research Council (ARC) through an Australian Laureate Fellowship awarded to JBH. 
We acknowledge the use of the National Computational Infrastructure (NCI) which is supported by the Australian Government, and accessed through the Sydney Informatics Hub (SIH) HPC Allocation Scheme 2022  (PI: TTG; CI: JBH).

LC acknowledges funding from the Chilean Agencia Nacional de Investigaci\'{o}n y Desarrollo (ANID) through Fondo Nacional de Desarrollo Cient\'{\i}fico y Tecnol\'{o}gico (FONDECYT) Regular Project 1210992.

This work presents results from the European Space Agency (ESA) space mission Gaia. Gaia data are being processed by the Gaia Data Processing and Analysis Consortium (DPAC). Funding for the DPAC is provided by national institutions, in particular the institutions participating in the Gaia MultiLateral Agreement (MLA). The Gaia mission website is https://www.cosmos.esa.int/gaia. The Gaia archive website is https://archives.esac.esa.int/gaia.

This work has used the following software products:
\href{http://www.starlink.ac.uk/topcat/}{TOPCAT}, \href{http://www.starlink.ac.uk/stil}{STIL}, and \href{http://www.starlink.ac.uk/stilts}{STILTS} \citep{2005ASPC..347...29T,2006ASPC..351..666T};
Matplotlib \citep{Hunter:2007};
IPython \citep{PER-GRA:2007};  
Astropy, a community-developed core Python package for Astronomy \citep{2018AJ....156..123A}; and Pynbody\footnote{\url{https://github.com/pynbody/pynbody}} 

\end{acknowledgements}

   \bibliographystyle{aa} 
   \bibliography{mybib,inprep,arxiv,dpac} 

\appendix
\section{Results from gas-dynamical simulations}
\label{sec:app}

Since classical Cepheid variable stars \citep{skowron2019ceph} are mostly younger than the rotation period at the Solar Circle (220 Myr), it is reasonable to assume that their distribution in phase space reflects the densest gas in the Galactic disc. Most of these will form in star clusters that disperse on the timescale of the disc rotation \citep{Bland-Hawthorn:2010}. The unbinding of the clusters injects about $\lesssim$5 km s$^{-1}$ of random motion into the phase space distribution \citep{mroz2019ceph}, well within the measurement uncertainty.
Therefore, the `Cepheid gap' can in principle reflect conditions in the gas from which these stars formed.

In order to investigate whether a bar or spiral arm resonance could lead to a similar gap in the distribution of gas in the disc, we ran a gas-dynamical/N-body simulation of an Milky Way barred surrogate. In brief, we approximate the Galaxy by a four-component model consisting of a DM halo, a stellar bulge, a stellar disc, and a cold ($10^3$ K), light ($\sim4\times10^9$ M$_\odot$) gas disc. The setup of the collisionless components is identical to the model discussed in \citet{Tepper-Garcia:2021}. The setup of the gas disc follows \citep{Tepper-Garcia:2022}. More details about this model will be provided elsewhere (Tepper-García et al, in prep.).

By virtue of the disc-to-total mass ratio of the model, the disc is subject to a bar instability. A bar forms after about 2.5 Gyr of evolution in isolation, and remains reasonably stable for at least another 1.5 Gyr.

To estimate the bar pattern speed $\Omega_p$ -- which dictates the location of the relevant resonances -- we rely on a Fourier analysis of the stellar disc's surface density. In brief, we calculate the amplitude ($A_2$) and the phase ($\phi_2$) of the \mbox{$m=2$} mode in a specified radial range ($R \leq 10$ kpc), and compute the phase change ($\dot{\phi}_2$) for the datum at which $A_2$ reaches a maximum, consistent with the estimate of the bar strength \citep[e.g.][]{Kataria2019}. As a check, we calculate both a pure arithmetic mean of and a radially-weighted average of $\dot{\phi}_2$ over the specified radial range \citep[e.g.][]{Athanassoula2003}. Each of this approaches yields a slightly different value \citep[see discusion in][]{Tepper-Garcia:2021}, but they are all consistent with $\Omega_p \approx 30 - 40$ \kmskpc{} for the bar at $T \approx 1.5$ Gyr after its formation.

\begin{figure}
    \centering
    \includegraphics[width=0.9\columnwidth]{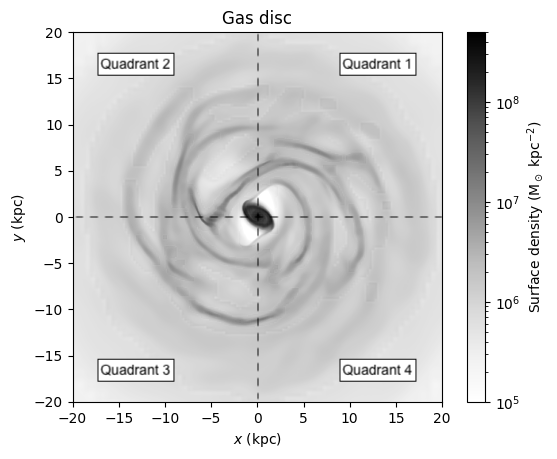}
    \caption{Distribution of gas in a Milky Way surrogate 1 Gyr after the formation of the central bar. For the purpose of the analysis, we divide the disc face into four quadrants identified by the numbers 1 through 4 as indicated in the figure. See also Fig.~\ref{fig:sims_Rvphi}.}
    \label{fig:sims_xy}
\end{figure}

\begin{figure*}
    \centering
    \includegraphics[width=0.45\textwidth]{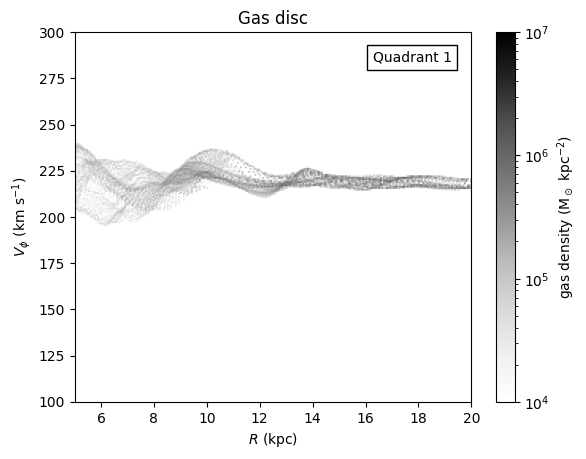}
    \includegraphics[width=0.45\textwidth]{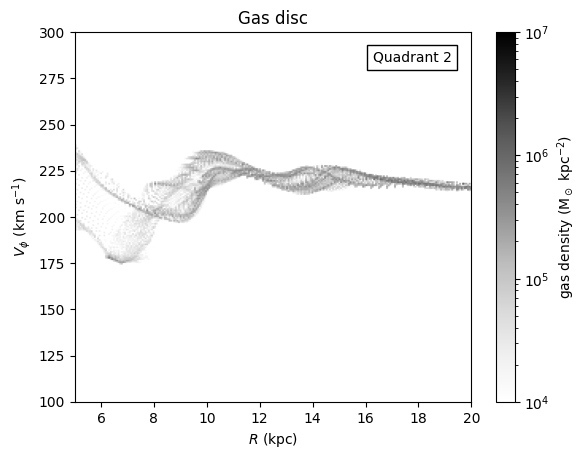}
    \includegraphics[width=0.45\textwidth]{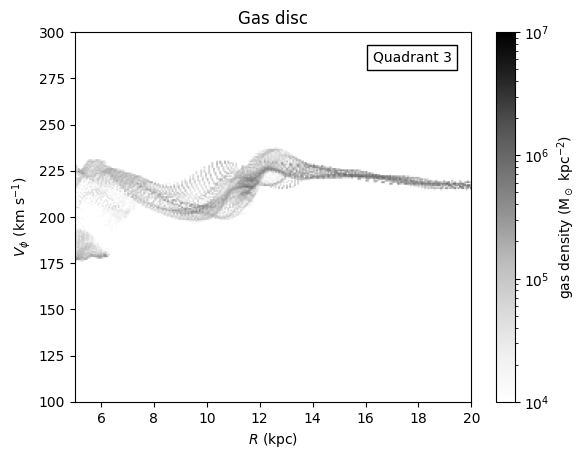}
    \includegraphics[width=0.45\textwidth]{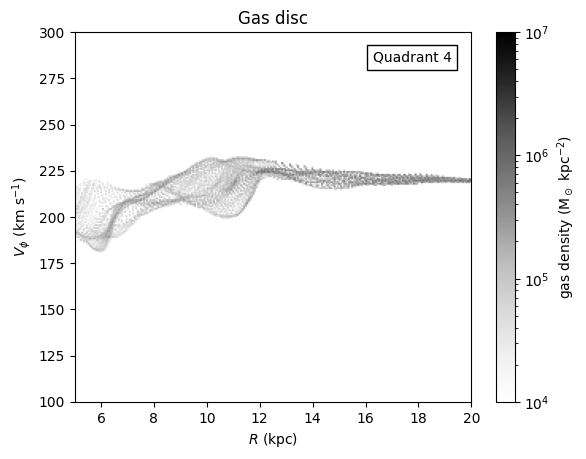}
    \caption{The rotation curve of the gas in each of the quadrants displayed in Fig.~\ref{fig:sims_xy} weighted by gas density. Note that no gap is apparent in any of the panels. See text for more details.}
    \label{fig:sims_Rvphi}
\end{figure*}

At this epoch, we looked at the distribution of gas in the synthetic galaxy at $T \approx 4$ Gyr (Fig.~\ref{fig:sims_xy}). We divided the volume occupied by the disc (i.e. the volume enclosed by $R = 20$ kpc and $|z| = 5$ kpc) into four quadrants, where `Quadrant 1' is defined by the region enclosed by $0 \leq \phi < 90$, `Quadrant 2' by $90 \leq \phi < 180$, and so on (where $\phi$ is the azimuthal angle), and analysed the distribution of gas in each of the quadrants in $R - V_\phi$ space, i.e. the rotation curve. The intent behind this is to see if the gas angular momentum ($L_Z$) undergoes any discontinuities, e.g. from disc resonances, gas compression and shocks, etc.
The result of this exercise is shown in Fig.~\ref{fig:sims_Rvphi}. In brief, none of the panels displays a clear gap in the distribution of gas, as seen in the rotation curve of the Cepheid population (Fig.~\ref{fig:azVel_and_Lz_vs_R}, left panel).

There may be several reasons for the disagreement. The most obvious one is our assumption that the dense gas in a thin 2D plane is an appropriate proxy for the Cepheid distribution. This is the most 
favourable situation for strong resonances to operate. But the resonances must set up on a very short timescale, i.e. less than one rotation period, and this is not easy to do.

Furthermore, it has long been known that the gas disc is corrugated with a wave amplitude of roughly 300 pc \citep[see references in][]{Tepper-Garcia:2022}. The phase-spiral effect is dominated by the younger stars \citep{Bland-Hawthorn:2019}. Another possibility is that the observed gap is transient, except that we produced a movie of the quadrant rotation curves, and no gaps were seen for one billion years after the bar formed. This needs to be revisited with higher resolution simulations.


Finally, it is possible that the gap is not triggered by the bar, and thus our isolated model is not able to reproduce it. This idea is supported by the fact that other recently discovered kinematic features in the Galaxy such as the phase spiral or the 11 kpc break reported by \citet{gaia_anticenter2021} are likely the result from a strong impulsive interaction  \citep[e.g.][]{Bland-Hawthorn:2021,McMillan:2022}.

\end{document}